\documentclass[preprint, showpacs,preprintnumbers,amsmath,amssymb]{revtex4}
\usepackage[dvips]{graphicx}
\usepackage{graphicx}
\usepackage{amsfonts}
\usepackage{bm}
\usepackage{amsmath}
\usepackage{amssymb}
\usepackage{color}
\usepackage[all]{xy}

\def\be{\begin{equation}}
\def\ee{\end{equation}}
\def\bea{\begin{eqnarray}}
\def\eea{\end{eqnarray}}

\begin{document}

\title{GEOMETRIC DESCRIPTION OF THE THERMODYNAMICS OF A BLACK HOLE WITH POWER
MAXWELL INVARIANT SOURCE}


\author{Gustavo Arciniega}
\email{gustavo.arciniega@nucleares.unam.mx} \affiliation{Instituto de
Ciencias Nucleares, Universidad Nacional Aut\'onoma de
M\'exico,\\A.P.
 70-543, M\'exico D.F. 04510 MEXICO}

\author{Alberto S\'anchez}
\email{asanchez@nucleares.unam.mx} \affiliation{Departamento de
posgrado, CIIDET,\\{\it AP752}, Quer\'etaro, QRO 76000, MEXICO}

\date{\today}

\begin{abstract}

Considering a nonlinear charged black hole as a thermodynamics
system, we study the geometric description of its phase
transitions.  Using the formalism of geometrothermodynamics we show
that the geometry of the space of thermodynamic equilibrium states
of this kind of black holes is related with information about
thermodynamic interaction, critical points and phase transitions
structure. Our results indicate that the equilibrium manifold of
this black hole is curved and that curvature singularities appear
exactly at those places where first and second order phase
transitions occur.


\end{abstract}


\maketitle

\section{Introduction}

AdS spacetimes in $n+1$--dimensions has won notoriety since the
norm/gravity correspondence was postulated in 1997 by Maldacena \cite{Maldacena,
Gubser, Witten} and black holes in the gravity side are related
with the temperature of the system in the field theory side. However the
 thermodynamic study of black holes in AdS spacetime was initiated long before the duality AdS/CFT appeared by
Hawking and Page \cite{Hawking} as a curiosity in theoretical physics. They show that the 
Schwarzschild-AdS black hole has a phase transition and this quality attracted others to keep studying this objects in AdS backgrounds. The Reissner-Nordstr\"om-AdS (RNAdS) black hole in $n+1$--dimensions
has been studied in \cite{Chamblin1} where they found that also RNAdS present a phase transition. Later, some authors considered the cosmological constant as a new thermodynamical variable in AdS black
holes \cite{Kastor, Dolan1}, extending in this way the phase space. 
Recently, nonlinear electrodynamics source in AdS-black holes rise interest in the community, but focusing in particular in the power Maxwell invariant field (PMI)
\cite{Hassaine, Hendi0, Maeda} where a $s$ parameter is taken as a power of the Maxwell term in the action, i.e. $(F_{\mu\nu}F^{\mu\nu})^{s}$, which reduces to the Maxwell field (linear electromagnetic
source) when $s=1$. In those theories the authors found that black holes with nonlinear source present a transition phase in both, the canonical ensemble and  the grand canonical ensemble. 

In order to describe the behaviour of the thermodynamic system by
means of geometry, different approaches have been proposed 
\cite{Weinhold, Ruppeiner, Ruppeiner2, quevedo2}
which use a Rimannian manifold to define a space of equilibrium
states where  the thermodynamic phenomena take place. These
approaches introduce a metric by means of which the thermodynamic
concepts are related to the geometry.

This work is organized as follows. In section \ref{gtd}, we present
the elements of the geometrothermodynamics formalism that will be
used in this work to obtain the thermodynamical behaviour of the
systems analyzed. In section \ref{termo}, we present the black hole
solution in AdS spacetime with power invariant Maxwell source
(PMI) and the thermodynamic attributes of this system. In section
\ref{secgtdPMI}, we apply the geometrothermodynamics (GTD)
formalism to black holes with PMI source. In the section \ref{RN},
we consider the case of the Reissner-Nordstr$\ddot{\mathrm{o}}$m
black hole where; its phase transitions are studied and we show that the
Legendre-invariant metric reproduces its corresponding phase
transition structure. Section \ref{Weinhold}, contains the results
of analyzing the thermodynamics of the black holes with PMI source
solution using the Weinhold geometry. In section
\ref{cosmological}, we analyze the thermodynamics behaviour of a
black hole with PMI source considering the cosmological constant
as a thermodynamical variable under GTD. Finally, in section
\ref{conclusions}, we present our conclusions.

\section{Geometrothermodynamics approach}\label{gtd}

Geometrothermodynamics (GTD) \cite{quevedo2, cesar} is a geometric formalism associated to physical systems in order to obtain its thermodynamical behaviour. The first element to consider under this formalism is a contact manifold $\mathcal{T}$ of dimension $(2n+1)$, called phase space, with a metric $G$ defined on it and a contact 1--form $\Theta$ that satisfies the condition  $\Theta \wedge (d\Theta)^n\neq 0$. On this geometrical structure, $(\mathcal{T}, G, \Theta)$, can be use a set of coordinates $\{Z^{A}\}_{A=1,\ldots,2n+1}\equiv\{\Phi,E^{a},I^{a}\}_{a=1,\ldots,n}$ and demand $G$ to be invariant by a total Legendre transformation. This means that when it is performing a change of coordinates  $\{Z^{A}\}_{A=1,\ldots,2n+1}$ to $\{\tilde{Z}^{A}\}_{A=1,\ldots,2n+1}$ that satisfies the transformations

\bea \label{gtd7} \Phi=\tilde{\Phi}-\tilde{E}_a
\tilde{I}^a\,,\quad \quad E^a=-\tilde{I}^a\,, \quad \quad
I^a=\tilde{E}^a\, , \eea

\noindent then  the new metric $\tilde{G}$ has the same structure as $G$ defined over its own set of coordinates. 

In GTD, we can also define an embedding $\varphi :\mathcal{E}\longrightarrow\mathcal{T}$ as $\varphi: (E^a) \longrightarrow (\Phi\,,E^a\,,I^a)$ with  $\Phi(E^a)$ such that $\varphi^*(\Theta)=\varphi^*(d\Phi-\delta_{ab}I^adE^b)=0$ holds. The $n$--dimensional Riemannian submanifold $\mathcal{E}$ is called the thermodynamic equilibrium space, $\Phi$ is called the thermodynamic potential, $E^{a}$ are called extensive variables and $I^{a}$ are called intensive variables. With these geometric elements, all the thermodynamic information of the system can be extracted by using the mapping $\varphi$ and the condition of the pullback acting on $\Theta$.

A GTD metric $G$ that satisfies the previous requirements and that we will be using in this work is given by

\bea \label{gtd6} G=\Theta^2+(\delta_{ab}E^a I^b)(\eta_{cd}dE^c
dI^d)\,,\eea where $\delta_{ab}={\mathrm{diag}}(1,1,\dots,1)$,
$\eta_{ab}={\mathrm{diag}}(-1,1,\dots,1)$ and $\Theta=d\Phi-\delta_{ab}I^adE^b$ and 
\bea \label{gtd4} \frac{\partial \Phi}{\partial E^a}=I_a\,, \quad
\quad  \quad \quad d\Phi= I_adE^a \,.\eea

We notice that equation $\Phi(E^{a})$ must be explicitly given and it corresponds to the fundamental equation in standard thermodynamics.

Applying the pullback $\varphi^*$ to the metric $G$, we get the induced metric $g$ by $\varphi^*(G)=g$ resulting in
\bea \label{gtd8} g^{GTD}=\varphi^*(G)=\Big( E^c
\frac{\partial \Phi}{\partial E^c} \Big)\Big(\eta_{ab}
\delta^{bc}\frac{\partial^2 \Phi}{\partial E^c
\partial E^d }dE^a dE^d\Big)\,.\eea 

We can see that it is only  necessary to know the fundamental equation $\Phi(E^{a})$ to determine explicitly the metric $g$ on $\mathcal{E}$. According to the GTD prescription, once we have $\Phi(E^{a})$, we have all the requirements to associate the thermodynamics of the physical system to this geometrical structure. In particular, our principal interest is to obtain the phase transitions of the system which should correspond to singular points in the curvature scalar $R$ as proposed in GTD \cite{quevedo1}. This procedure to obtain the phase transitions of black holes systems with the GTD approach has been used successfully in previous works \cite{antonio, Bravetti, quevedo3, quevedo4, Aman, Aman2, Aman3}. 

\section{Thermodynamics}
\label{termo}

The corresponding line element for the black hole with power
Maxwell invariant (PMI) source \cite{Hendi} solution is,

\bea \label{metric} ds^2=-f(r)dt^2+\frac{dr}{f(r)}+r^2
d\Omega^2{}_{d-2}\,,\eea where $d\Omega^2{}_{d-2}$ stands for
standard element on $S^d$, with lapse function,

\bea \label{equ11} f(r)=1+\frac{r_+^2}{l^2}-\frac{m}{r_+{}^{n-2}}
+\frac{(2s-1)^2\Big[\frac{(n-1)(2s-n)^2 q^2}{(n-2)(2s-1)^2}
\Big]^s}{(n-1)(n-2s) r_+{}^{\frac{2(ns-3s+1)}{2s-1}}}\,.\eea Here
$m$ and $q$ are related with the ADM mass $M$ and the electric
charge $Q$ by means of the relation,

\bea \label{equ31} m&=&\frac{16\pi M }{(n-1)\omega_{n-1}}\,,\\
\label{equ312} q&=&\Bigg[\frac{8\pi}{\sqrt{2} s
\omega_{n-1}}\Bigg]^{\frac{1}{2s-1}}\Bigg[\frac{n-2}{n-1}
\Bigg]^{\frac{1}{2}}\frac{(2s-1)^{\frac{2s-2}{2s-1}}}{n-2s}Q^{\frac{1}{2s-1}}\,.\eea

The horizons of the black hole with PMI source metric correspond
to the roots of the lapse function $f(r)$. In terms of the
exterior horizon radius $r_+$, the black hole mass is given by the
expression,

\bea \label{equ22} M(r_+,Q)&=&\frac{(n-1)\omega_{n-1}
}{16\pi}\Bigg[r_+{}^{n-2}+\frac{r_+{}^n}{l^2}-\frac{(2s-1)^{2-2s}(n-1)^{s-1}(2s-n)^{2s-1}}{(n-2)^s}r_+{}^{\frac{(2s-n)}{2s-1}}q^{2s}\Bigg]\,.\nonumber\eea

Furthermore, the entropy of the black hole with PMI source  is
defined as

\bea \label{equ3} S=\frac{\omega_{n-1} r_+{}^{n-1}}{4}\,,\eea

In terms of this entropy, the corresponding thermodynamic
fundamental equation is given  by

\bea \label{equ5} M(S,Q)=\frac{(n-1)\omega_{n-1}
}{16\pi}\Bigg\{&&\Bigg[
\frac{4S}{\omega_{n-1}}\Bigg]^{\frac{n-2}{n-1}}+\Bigg[
\frac{4S}{\omega_{n-1}}\Bigg]^{\frac{n}{n-1}}l^{-2}-\nonumber
\\&-&\frac{(2s-1)^{2-2s}(n-1)^{s-1}(2s-n)^{2s-1}}{(n-2)^s}\Bigg[
\frac{4S}{\omega_{n-1}}\Bigg]^{\frac{2s-n}{(n-1)(2s-1)}}q^{2s}\Bigg\}\,,\nonumber
\\ \eea with $ \omega_{n-1}=(2\pi^{n/2})/\Gamma(n/2)$. In order to avoid inconsistent results, we
will consider that the parameter $s$ satisfies the
relationships $2s-1=i$ and $2s-n=i-n+1$, with $i$ any integer and
$n\neq i+1$ in order to keep real all terms in equation (\ref{equ22}). The physical parameters of the black hole with PMI
source satisfy the first law of black hole thermodynamics
\cite{davies},

\bea \label{flaw} dM=TdS+\Phi dQ\,,\eea where $T$ is the Hawking
temperature which is proportional to the surface gravity on the
horizon and  $\Phi$ the electrical potential. As in ordinary
thermodynamics, all the thermodynamic information is contained in
the fundamental equation. The standard conditions of the
thermodynamic equilibrium are given by the expressions,

\bea \label{c2} T=\frac{\partial M}{\partial S}\,,\quad \quad 
\quad  \quad \Phi=\frac{\partial M}{\partial Q}\,.\eea

According with the Ehrenfest's classification \cite{callen} the
phase transitions take place where the derivatives of the Gibbs
free energy $G=T\,M-\Phi Q$ diverge, for example,  the first-order
phase transitions exhibit a discontinuity in the first derivative
of the $G$ with respect to some thermodynamic variable and
second-order phase transitions exhibit discontinuity in a second
derivative of $G$. Because the heat capacity at constant electric
charge $C_Q$ is a second derivative of the free energy $G$, it
will be used to describe the second order phase transition
structure of the system,

\bea \label{c3} C_Q=- T\Bigg(\frac{\partial^2 G}{\partial T^2}
\Bigg)_Q=T\Bigg(\frac{\partial S}{\partial T}
\Bigg)_Q=\frac{\Bigg(\frac{\partial M}{\partial S}
\Bigg)_Q}{\Bigg(\frac{\partial^2 M}{\partial S^2} \Bigg)_Q}\,.\eea

Using the equation (\ref{equ5}) it is possible to compute the heat
capacity at constant electric charge. However, it is not possible
to write this quantity in a compact form. Therefore, we will take particular configurations of the black hole
and its behaviour using different values of the
parameter $s$ and $n$. As an example of this analysis we consider
the values $s=\frac{5}{2}$ and $n=4$, corresponding to $i=4$. The $C_Q$ has the form,

\bea \label{c3-1} C_Q=\frac{6S\Bigg[300 S^{\frac{5}{4}}+2\cdot
2^{\frac{3}{8} }S^{\frac{3}{4}} \pi^{\frac{5}{4}}l^2
Q^{\frac{5}{4}}+75\cdot 2^{\frac{1}{3} }\pi^{\frac{4}{3}} l^2
S^{\frac{7}{12}}\Bigg]}{600 S^{\frac{5}{4}}-11\cdot 2^{\frac{3}{8}
}5^{\frac{3}{4}} \pi^{\frac{5}{4}}l^2 Q^{\frac{5}{4}}-150\cdot
2^{\frac{1}{3} }\pi^{\frac{4}{3}} l^2 S^{\frac{7}{12}}}\,.\eea

\begin{figure}[h]
{\includegraphics[width=10cm]{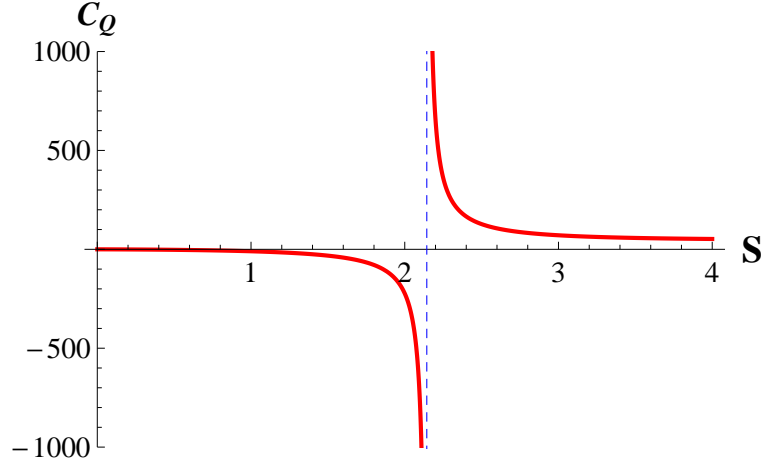} \caption{The heat
capacity $C_{Q}$ as a function of the entropy $S$, for $l=1$, 
$Q=1$, $s=5/2$ and $n=4$.}}\end{figure}

The second order phase transitions take place at those points
where the heat capacity diverges, i. e.

\bea \label{c3-2} 600 S^{\frac{5}{4}}-11\cdot 2^{\frac{3}{8}
}5^{\frac{3}{4}} \pi^{\frac{5}{4}}l^2 Q^{\frac{5}{4}}-150\cdot
2^{\frac{1}{3} }\pi^{\frac{4}{3}} l^2 S^{\frac{7}{12}}=0\,.\eea
The  behaviour of $C_Q$ is depicted in figure 1. We can see that $C_Q$
has points of divergence which tell us, according with Ehrenfest's
theory, that there is a  point where a second order phase
transition can take place.

Using the conditions of the thermodynamic equilibrium (\ref{c2}),
one obtains the temperature and electric potential of the black
hole on the event horizon as

\begin{center}
\bea \label{temperature}
T&=&\frac{\pi^{\frac{5}{3}}}{2^{\frac{1}{6}}S^{\frac{11}{12}}l^2}\Bigg[300
S^{\frac{5}{4}}+2\cdot 2^{\frac{3}{8} }5^{\frac{3}{4}}
\pi^{\frac{5}{4}}l^2 Q^{\frac{5}{4}}+75\cdot 2^{\frac{1}{3}
}\pi^{\frac{4}{3}} l^2 S^{\frac{7}{12}} \Bigg]\,, \\  \Phi &=&
\frac{2^{\frac{17}{24}}5^{\frac{3}{4}}}{10\pi^{\frac{5}{12}}}S^{\frac{1}{12}}Q^{\frac{1}{4}}\,.
\eea\end{center}

We see that all intensive thermodynamic variables are
well-behaved. The general behaviour of these variables are
illustrated in figure 2.

\begin{center}
\begin{figure}[h]
{\includegraphics[width=7cm]{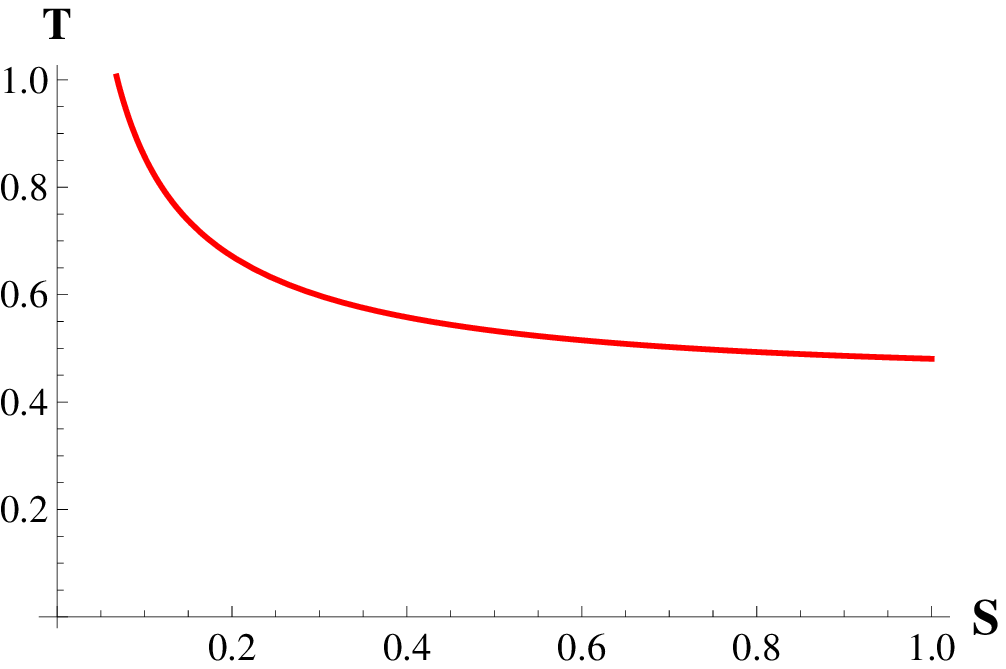}
\includegraphics[width=7cm]{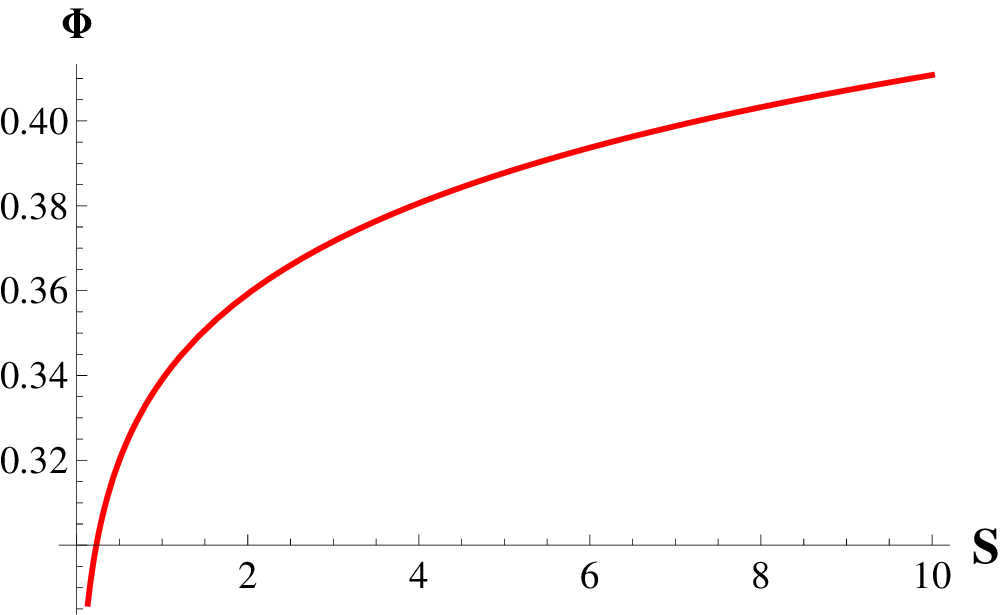}\caption{The temperature $T$ (left) and the electric potential $\Phi$ (right) as  functions of the entropy $S$, for $l=1$ and
$Q=1$. In  these graphics we also consider: $s=5/2$ and $n=4$} }\end{figure}\end{center}

In the same way we can analyze other examples, and in all of them we
found a similar behaviour, i.e., there are points of divergence
that corresponds to second-order phase transitions. Other examples
are shown in the figure 3.

\begin{center}
\begin{figure}[h]
{\includegraphics[width=7cm]{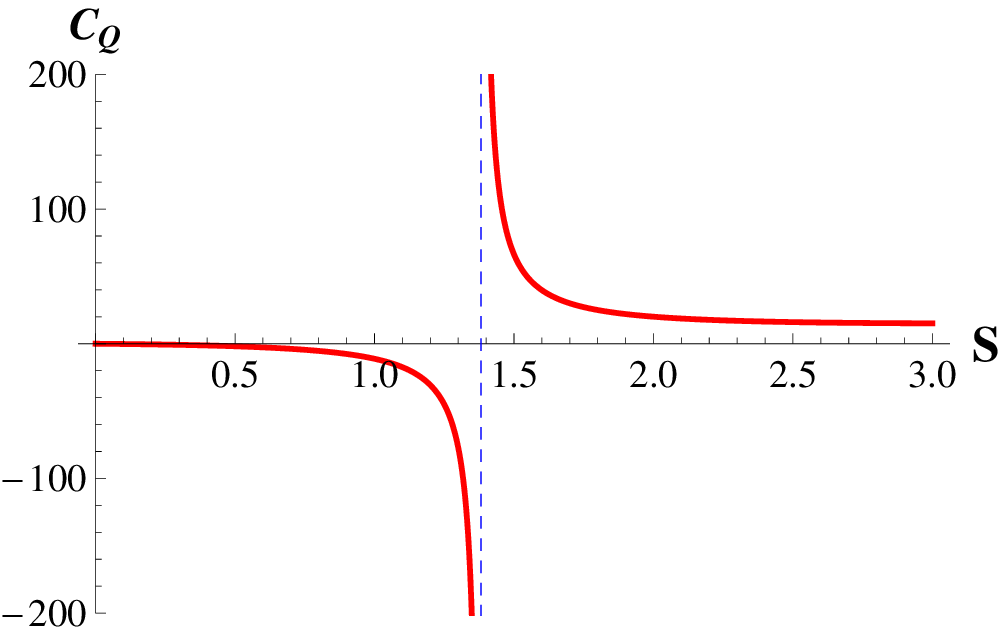}
\includegraphics[width=7cm]{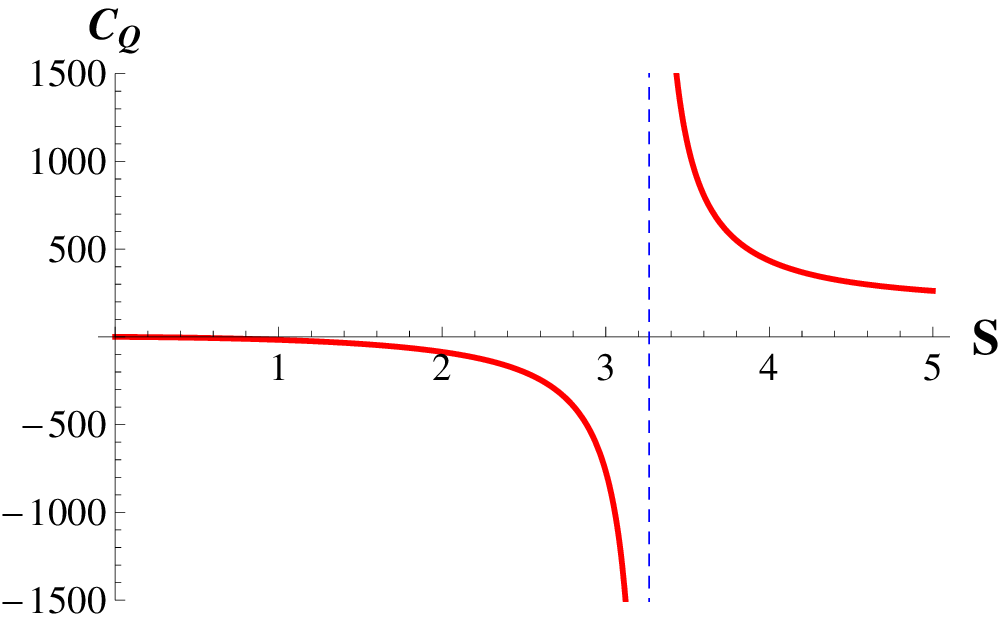}\caption{The heat
capacity $C_{Q}$ as a function of the entropy $S$, for $l=1$ and
$Q=1$. In  these graphics we also consider: $s=5/2$ and $n=3$ (left), $n=6$ (right)} }\end{figure}\end{center}

\newpage

\section{Geometrothermodynamics of the black hole with PMI source }
\label{secgtdPMI}

The metric (\ref{gtd8}) with $\Phi=M$ and $E^a=\{S,Q\}$
reads,

\bea \label{GTDmetric1} g^{GTD}=\Big( S \frac{\partial
M}{\partial S} + Q \frac{\partial M}{\partial
Q}\Big)\Big(-\frac{\partial^2 M}{\partial S^2 }dS^2
+\frac{\partial^2 M}{\partial Q^2 }dQ^2\Big) \,.\eea

Using the expressions for the $M$, as was given in Eq.
(\ref{equ5}), we compute the curvature scalar $ R{}^{GTD}$
corresponding to the metric (\ref{GTDmetric1}). We found that the corresponding
curvature scalar $R{}^{GTD}$ is  different to zero. This result indicates, in accordance with 
GTD, that in this black hole there exists thermodynamic interaction. We don't write the explicit form of the curvature scalar because it is a cumbersome expression that cannot be written in a compact form. However taking the same values used in the heat capacity
(\ref{c3-1}), we can write the curvature scalar as,

\bea \label{scalarpart}
R{}^{GTD}=\frac{\mathcal{F}(S,Q)}{\mathcal{P}_1{}^3
\mathcal{P}_2{}^2}\,,\eea with,

\bea \label{pol1} \mathcal{P}_1&=&300 S^{\frac{5}{4}}+32\cdot
2^{\frac{3}{8} }5^{\frac{3}{4}} \pi^{\frac{5}{4}}l^2
Q^{\frac{5}{4}}+75\cdot 2^{\frac{1}{3} }\pi^{\frac{4}{3}} l^2
S^{\frac{7}{12}}\,,\\ \label{pol2} \mathcal{P}_2 &=& 600
S^{\frac{5}{4}}-11\cdot 2^{\frac{3}{8} }5^{\frac{3}{4}}
\pi^{\frac{5}{4}}l^2 Q^{\frac{5}{4}}-150\cdot 2^{\frac{1}{3}
}\pi^{\frac{4}{3}} l^2 S^{\frac{7}{12}}\,,\eea so the
curvature singularities are determined by the zeros of the two
polynomials entering the denominator. The function $\mathcal{F}(S,
Q )$ is a polynomial that is different from zero at those points
where the denominator vanishes. It follows that there exist
curvature singularities at those points where the condition $600
S^{\frac{5}{4}}-11\cdot 2^{\frac{3}{8} }5^{\frac{3}{4}}
\pi^{\frac{5}{4}}l^2 Q^{\frac{5}{4}}-150\cdot 2^{\frac{1}{3}
}\pi^{\frac{4}{3}} l^2 S^{\frac{7}{12}}=0 $ is satisfied. These
singularities coincide with the points where $C_Q\longrightarrow
\infty$, i. e., with the points where second order phase
transitions take place, as given in equation (\ref{c3-2}). The general
behaviour of the curvature scalar is illustrated in figure 4.

\begin{center}
\begin{figure}[h]
{\includegraphics[width=10cm]{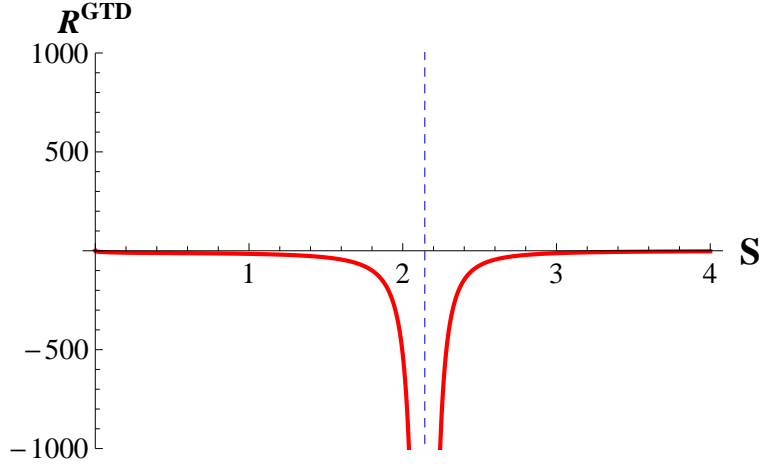} \caption{The
curvature scalar $R^{GTD} $ as a function of the entropy $S$, with
$l=1$, $Q=1$, $s=5/2$ and $n=4$.}}\end{figure}
\end{center}

The behaviour of the curvature scalar, corresponding
to the same values used in the heat capacities shown in figure 2,
is shown in the figure 5.

\begin{center}
\begin{figure}[h]
{\includegraphics[width=7cm]{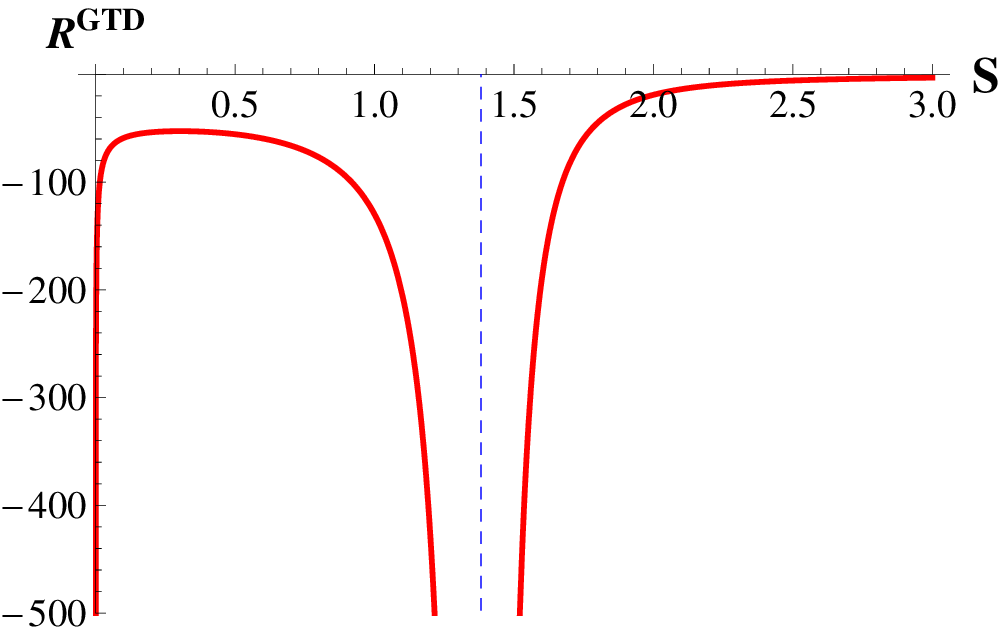}
\includegraphics[width=7cm]{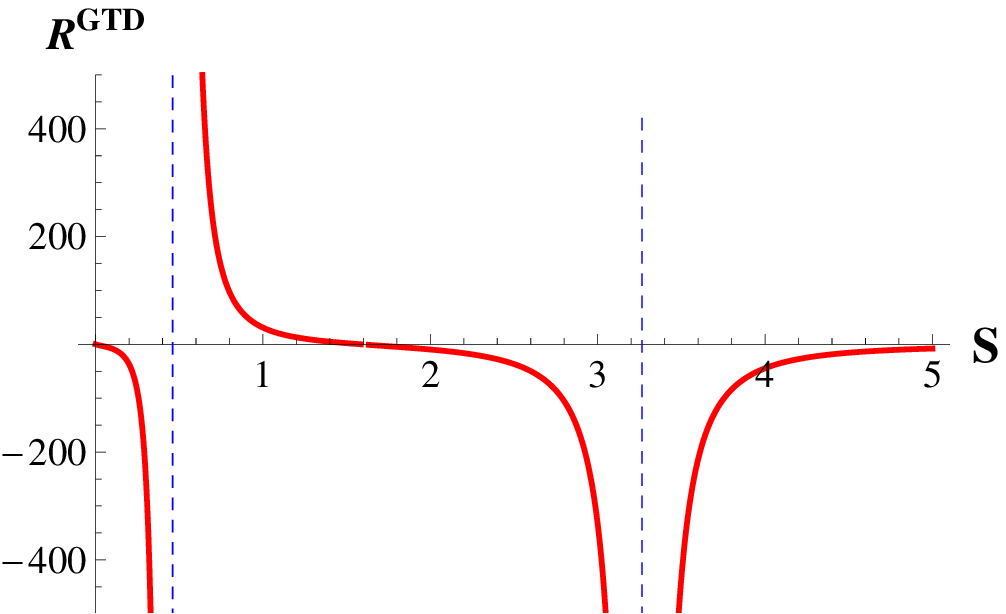}\caption{The curvature scalar $R^{GTD}$
as a function of the entropy $S$, for $l=1$ and
$Q=1$. In  these graphics we also consider: $s=5/2$ and $n=3$ (left), $n=6$ (right)} }\end{figure}
\end{center}

As we can see the curvature scalar corresponding to $n=6$ have two
singularities; one of them,  which is between zero and one,
corresponds to $f=S \frac{\partial M}{\partial S} + Q
\frac{\partial M}{\partial Q}=0$ as is shown in figure 6; therefore, we should not
consider it as a physical singularity. The other one, corresponds to
the point where the capacity becomes infinite, which once again
shows the correspondence between curvature singularities and
second order phase transitions. The same happens for other
examples

\begin{center}
\begin{figure}[h]
{\includegraphics[width=10cm]{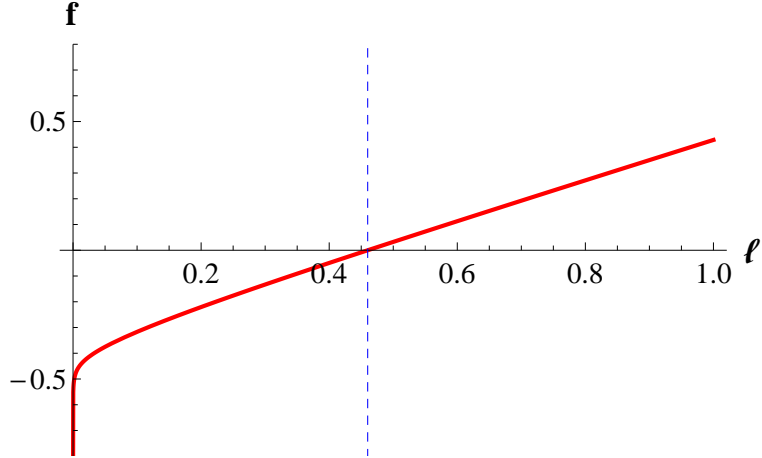} \caption{The factor
$f= S \frac{\partial M}{\partial S} + Q \frac{\partial M}{\partial
Q}$ as a function of $S$, with $Q=1$, $l=1$, $s=5/2$ and
$n=6$.}}\end{figure}
\end{center}

These results tell us, according with  geometrothermodynamics,
that there exist curvature singularities at those points where
second order phase transitions occur \cite{quevedo1}.

As a result we find all the intervals where the curvature scalar
becomes singular correspond to divergent points of the heat
capacity; so, geometrothermodynamics correctly describes the second
order phase transitions of this thermodynamic system.

\section{The case $s=1$ }
\label{RN}

In $(n + 1)$--dimensions the PMI theory becomes conformally
invariant, the same as Maxwell theory in four-dimensions, if the
parameter has the value  $s = (n + 1)/4$ \cite{Hassaine}. When we
consider the particular case $s=1$, we return to the  Maxwell
theory, i.e., a Reissner-Nordstr$\ddot{\mathrm{o}}$m black hole with linear electromagnetic source.
In this section we will analyze this case in order to continue
showing the consistency of GTD.

If $s=1$, then $n$ can be equal to three or larger. We choose $n=3$ and the  fundamental
equation takes the form,

\bea \label{equ55} M(S,Q)=\frac{2\omega_{2} }{16\pi}\Bigg\{\Bigg[
\frac{4S}{\omega_{2}}\Bigg]^{\frac{1}{2}}+\Bigg[
\frac{4S}{\omega_{2}}\Bigg]^{\frac{3}{2}}l^{-2}+\Bigg[
\frac{4S}{\omega_{2}}\Bigg]^{-\frac{1}{(2)}}q^{2}\Bigg\}\,,\nonumber
\\ \eea with $ \omega_{2}=(2\pi^{3/2})/\Gamma(3/2)$ and $q=(8\pi Q)/ \omega_{2}$. Using the
equation (\ref{equ55}) into the metric (\ref{GTDmetric1}) we get
the corresponding curvature scalar $R{}^{GTD}$,

\bea \label{scalarRN}
R{}^{GTD}=\frac{\mathcal{G}(S,Q)}{\Big(l^2\pi
S+3S^2+3\pi^2Q^2l^2\Big)^3 \Big( 3S^2-\pi Sl^2+3\pi^2 Q^2
l^2\Big)^2}\,,\eea with,

\bea \label{GRN} \mathcal{G}(S,Q)=48\pi^3S^2 l^4\Big[&&
45S^5-6\pi^4SQ^2l^6+15\pi^3S^2Q^2 l^4+6\pi^5Q^4 l^6-18\pi^2S^3
 Q^2 l^2- \nonumber \\ &-&63\pi^4 S Q^4 l^4-2\pi^2 S^3 l^4-3\pi S^4 l^2\Big]\,.\eea

As we can see $R^{GTD}\neq 0$ indicating thermodynamic
interaction. The heat capacity at constant electric charge is
computed by means of the equation (\ref{c3}):

\bea \label{cRN} C_Q=\frac{2S\Big(3S^2+\pi S l^2-\pi^2Q^2 l^2
\Big)}{3S^2-\pi Sl^2+3\pi^2 Q^2 l^2}\,.\eea

We see from the expression for the curvature scalar that the
singular points correspond to the values that fulfill the equation
$3S^2-\pi Sl^2+3\pi^2 Q^2 l^2=0$ which are exactly the points where
a phase transition occurs in the heat capacity (\ref{cRN}). This behaviour is shown in figure 7.

\begin{center}
\begin{figure}[h]
{\includegraphics[width=7cm]{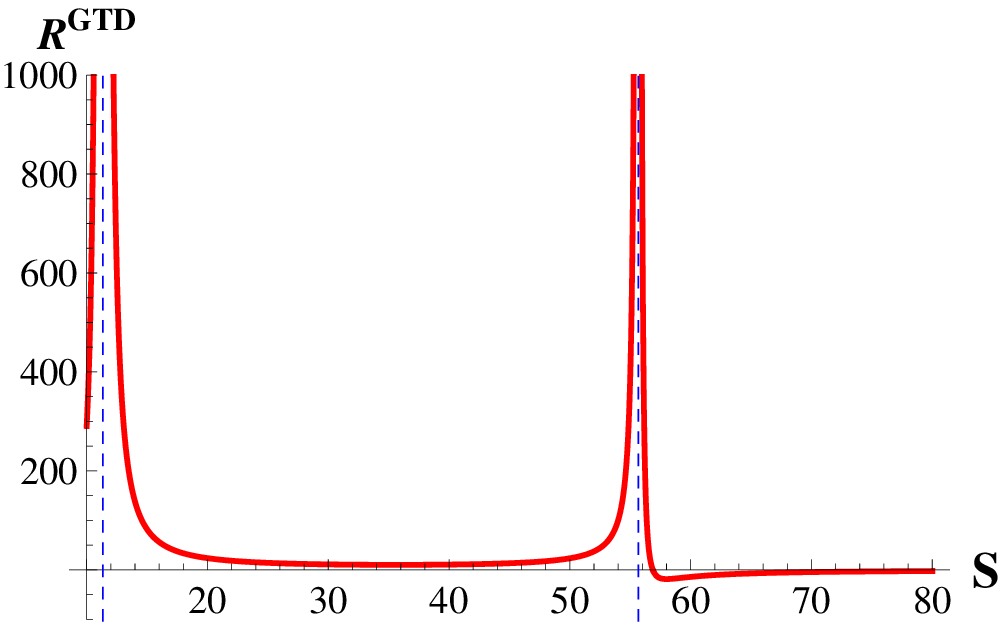}
\includegraphics[width=7cm]{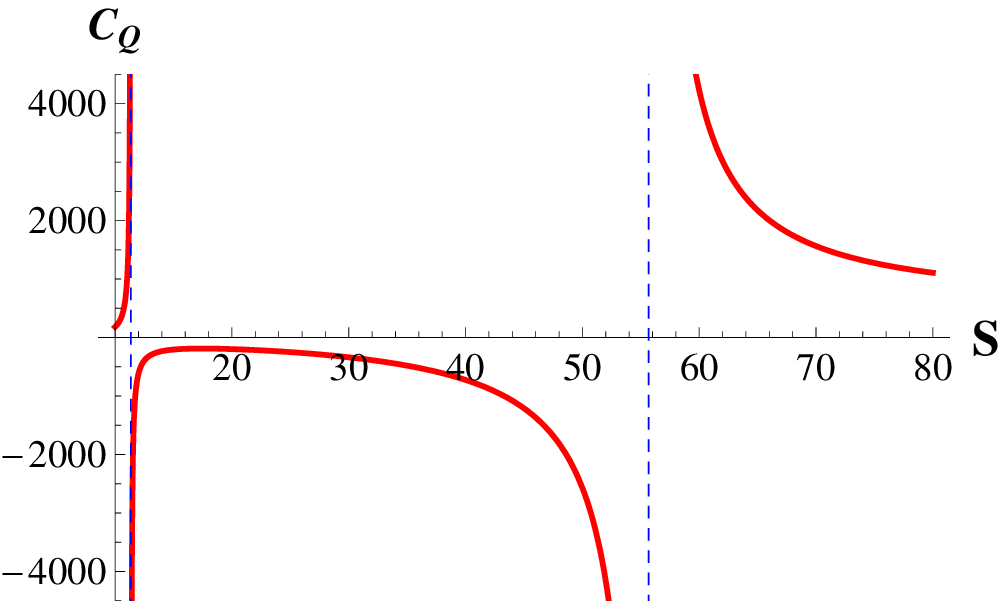}\caption{Curvature scalar $R^{GTD}$ (left) and heat capacity (right)
as  functions of  $S$, for $Q=1$, $l=8$, $s=1$ and $n=3$.} }\end{figure}
\end{center}

We can notice that this RNAdS black hole has two points where heat capacity becomes singular and it is worthwhile to mention that these are the same points where the curvature scalar diverges. According with Ehrenfest classification and the GTD prescription this tell us that this black hole has two second order phase transition in agreement with \cite{quevedo3,Liu}.

\newpage

\section{The Weinhold approach }
\label{Weinhold}

In this section we analyze the thermodynamic geometry of the black
hole with PMI source by using the Weinhold metric which is defined
as \cite{Weinhold}

\bea \label{weinmetric1} g^{W}= \frac{\partial^2 M}{\partial S^2
}dS^2 + 2\frac{\partial^2 M}{\partial S \partial Q }dS dQ+
\frac{\partial^2 M}{\partial Q^2 }dQ^2 \,.\eea

Using the mass $M$ of the black hole with PMI source (\ref{equ5})
the metric (\ref{weinmetric1}) can be calculated. The metric and
corresponding scalar curvature are very complicated so we will not
present them here. Instead we present one of the particular
examples studied in the section (\ref{secgtdPMI}). Considering
$s=\frac{5}{2}$ and $n=4$ the curvatures scalar can be written as

\bea \label{scalarwein} R^{W}=\frac{1500\cdot 2^{\frac{1}{12}}
\cdot 5^{\frac{1}{2}} \pi^{\frac{9}{2}} S^2 l^4}{\Big[15\cdot
2^{\frac{3}{8}} \cdot 5^{\frac{3}{4}} \pi^{\frac{25}{12}}
S^{\frac{4}{3}} l^2-60\cdot 2^{\frac{1}{24}} \cdot 5^{\frac{3}{4}}
\pi^{\frac{3}{4}} S^2+8\cdot 2^{\frac{5}{12}} \cdot
5^{\frac{1}{2}} \pi^2 S^{\frac{3}{4}} Q^{\frac{5}{4}} l^2
\Big]^2}\,,\eea

The heat capacity at constant electric charge is given by the
expression (\ref{c3}). We can see that the Weinhold geometry is
curved, signaling interaction for this thermodynamic system. There
are singular points, but they are not consistent with the ones of
the heat capacity for fixed charge. In figure 8 we plot the
scalar curvature and the heat capacity and in figure 9 we plot both functions in order to
illustrate this result.

\begin{figure}[h]
{\includegraphics[width=7cm]{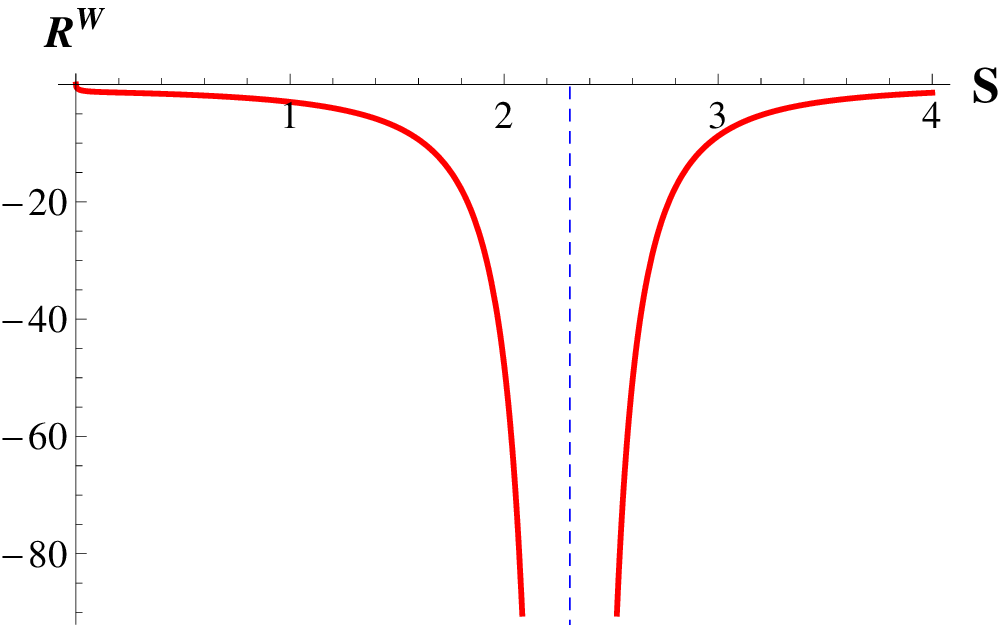}
\includegraphics[width=7cm]{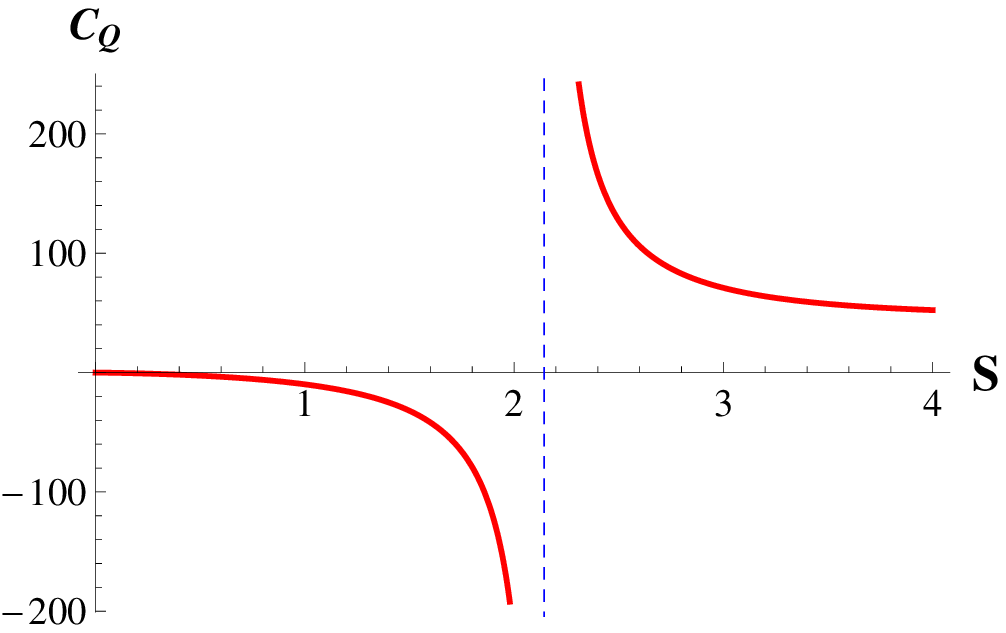}\caption{The
curvature scalar $R^W$ (left) and the heat capacity (right) as functions of $S$, with $Q=8$, $s=5/2$ and
$n=4$.}}\end{figure}

\begin{figure}[h]
{\includegraphics[width=10cm]{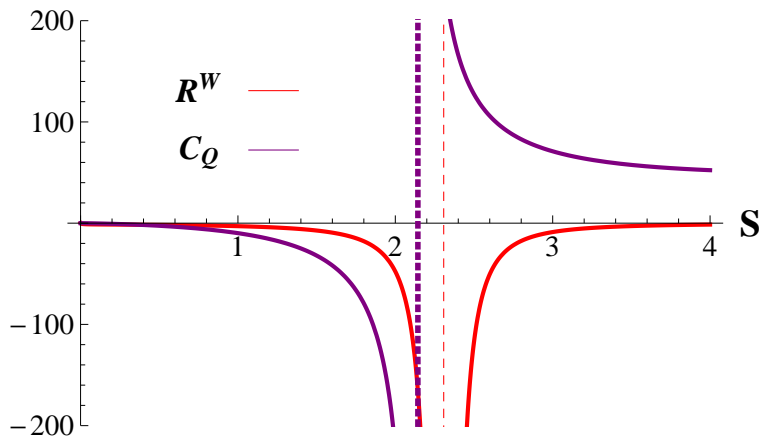}
\caption{The
curvature scalar $R^W$ (in red) and  heat capacity (in purple) as functions of $S$, with $Q=8$, $s=5/2$ and
$n=4$. The dotted thick vertical line corresponds to the point where the heat capacity is singular and the dashed vertical line is where the curvature scalar of the Weinhold metric diverges.}}\end{figure}

The fact that this Weinhold curvature scalar doesn't diverge at the same points where the heat capacity does, tell us that it is not possible to associate curvature singularities with second-order phase transitions.

The case of Ruppeiner's geometry  for thermodynamical systems \cite{Ruppeiner, Ruppeiner2} must be computed in the entropy representation, but this AdS black hole solution with nonlinear electromagnetic source doesn't allow us to write an explicit expression for the entropy in terms of the remaining variables. However, as has been shown in \cite{quevedoBH, Mrugala}, one can prove that Ruppeiner's metric, $g^{R}$, is proportional to Weinhold's metric, $g^{W}$ as $g^{R}=(1/T)g^{W}$, where $T$ is the temperature. This is why we expect that Ruppeiner's approach exhibits an equivalent behaviour as Weinhold's approach. 

\section{Cosmological constant as a thermodynamic variable }
\label{cosmological}

Next, we will consider $l$ as a thermodynamic variable \cite{hernando5}.  The first law of thermodynamics can be written as

\bea \label{flaw1} dM=TdS+\Phi dQ+L dl\,,\eea where
$L=\frac{\partial M}{\partial l}$ is the thermodynamic variable
dual to $l$, the cosmological constant. Then, by means of relationship for the mass
(\ref{equ5}) we can obtain intensive thermodynamics variables
$T$, $\Phi$ and $L$. Unfortunately, it is not possible to write
these quantities in a simple form. In order to solve this problem,
as we have done in the section (\ref{termo}), we will consider some configurations of the black hole taking particular values for the different parameters. Taking the values: $s=\frac{5}{2}$ and $n=4$,
corresponding to $i=4$, the  $C_Q$ has the form

\bea \label{c3-11} C_Q=\frac{6\Bigg[75\cdot
2^{\frac{1}{6}}\pi^{\frac{25}{12}} S^{\frac{7}{12}}l^2+150\cdot
2^{\frac{5}{6} }\pi^{\frac{3}{4}} S^{\frac{5}{4}}
+2^{\frac{29}{24} }\cdot 5^{\frac{3}{4}}\pi^{2}
Q^{\frac{5}{4}}l^2\Bigg]}{300\cdot2^{\frac{5}{6}
}\pi^{\frac{3}{4}}S^{\frac{5}{4}}-11\cdot 2^{\frac{5}{24} }\cdot
5^{\frac{3}{4}} \pi^{2} Q^{\frac{5}{4}}l^2-150\cdot 2^{\frac{1}{6}
}\pi^{\frac{25}{12}} S^{\frac{7}{12}}l^2}\,.\eea

\begin{center}
\begin{figure}[h]
{\includegraphics[width=10cm]{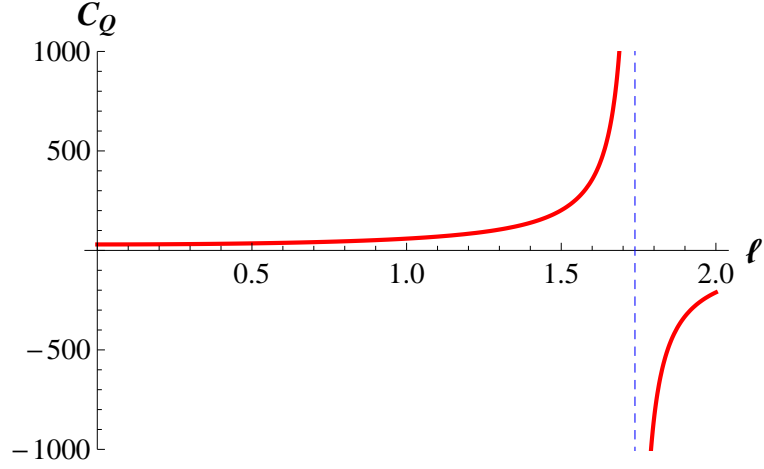} \caption{The
heat capacity $C_{Q}$ as a function of  $l$, for $Q=1$, 
$S=10$, $s=5/2$ and $n=4$.}}\end{figure}
\end{center}

The second order phase transitions take place at those points
where the heat capacity diverges, i. e., for

\bea \label{c3-22} 300\cdot2^{\frac{5}{6}
}\pi^{\frac{3}{4}}S^{\frac{5}{4}}-11\cdot 2^{\frac{5}{24} }\cdot
5^{\frac{3}{4}} \pi^{2} Q^{\frac{5}{4}}l^2-150\cdot 2^{\frac{1}{6}
}\pi^{\frac{25}{12}} S^{\frac{7}{12}}l^2=0\,.\eea The  behaviour of
$C_Q$ is depicted in figure 10. Using the conditions of the
thermodynamic equilibrium we get the intensive thermodynamic
variables:

\bea \label{temperature1} T&=&\frac{2^{\frac{1}{2}}}{300
\pi^{\frac{29}{12}} S^{\frac{11}{12}}l^2}\Bigg[75\cdot
2^{\frac{1}{6}}\pi^{\frac{25}{12}} S^{\frac{7}{12}}l^2+150\cdot
2^{\frac{5}{6} }\pi^{\frac{3}{4}} S^{\frac{5}{4}}
+2^{\frac{29}{24} }\cdot 5^{\frac{3}{4}}\pi^{2} Q^{\frac{5}{4}}l^2
\Bigg]\,, \\  \Phi &=& \frac{2^{\frac{17}{24}}\cdot5^{\frac{3}{4}}
}{10\pi^{\frac{5}{12}}}S^{\frac{1}{12}}Q^{\frac{1}{4}}\,, \\
L&=&-\frac{3\cdot
2^{\frac{1}{3}}}{2\cdot\pi^{\frac{5}{3}}}\frac{S^{\frac{4}{3}}}{l^2}\,.
\eea

We see that all intensive thermodynamic variables are
well-behaved. The general behaviour of these variables are
illustrated in figure 11.

\begin{figure}[h]
{\includegraphics[width=6.5cm]{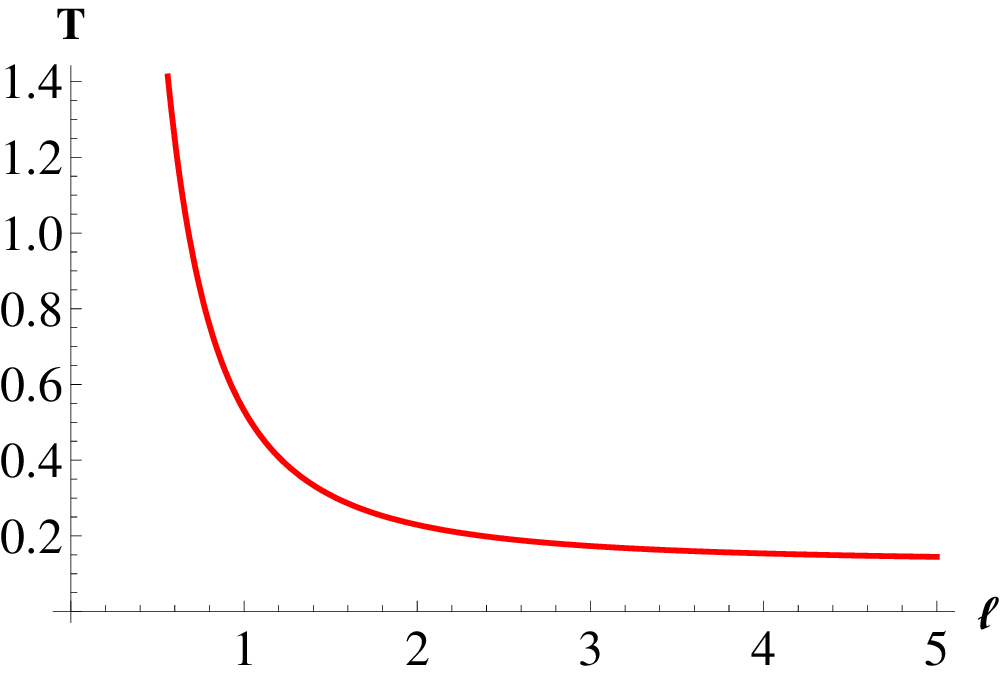}
\includegraphics[width=6.5cm]{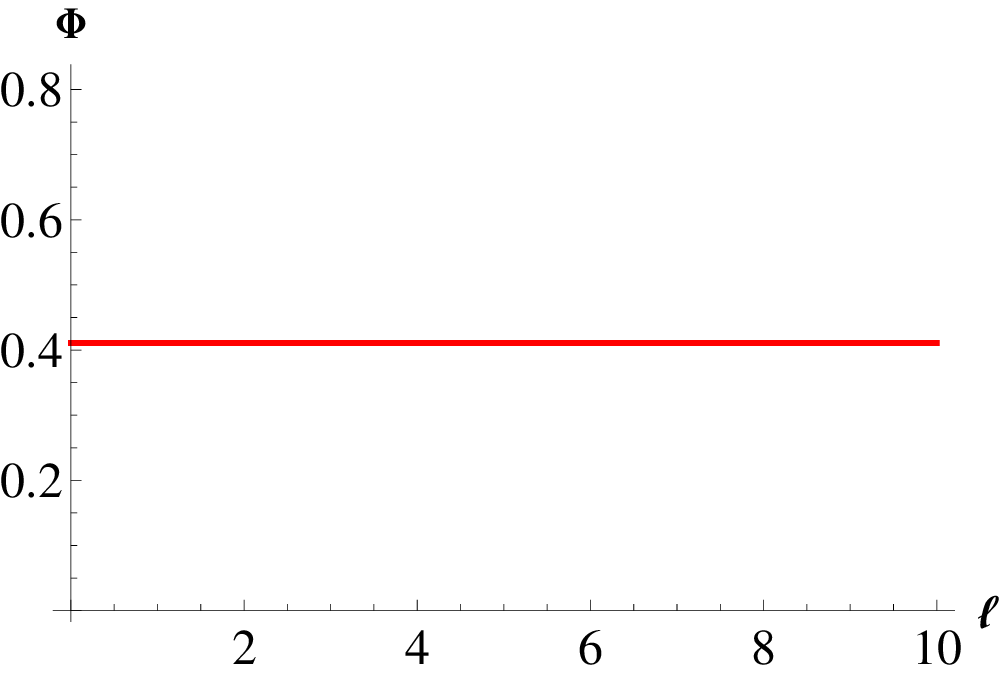}\includegraphics[width=6.5cm]{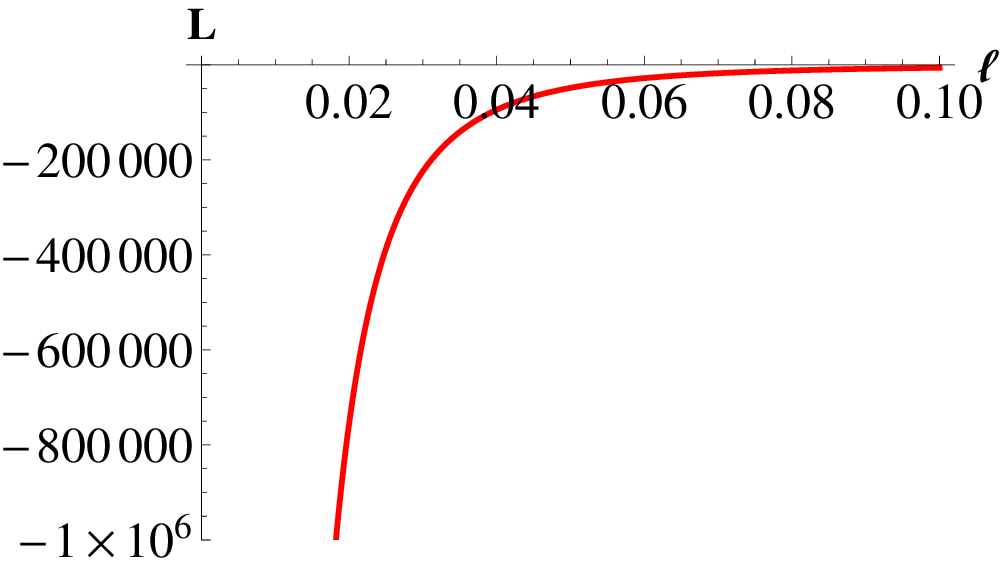}\caption{$T$ (left), $\Phi$ (center) and $L$(right)
as  functions of the $l$, for $Q=1$, 
$S=10$, $s=5/2$ and $n=4$.}  }\end{figure}

Then, using the relationship for the induced metric $g$ (\ref{gtd8})  with $\Phi=M$ and
$E^a=\{S,Q,l\}$ we get,

\bea \label{GTDmetric11} g_I^{GTD}=\Big( S \frac{\partial
M}{\partial S} + Q \frac{\partial M}{\partial Q}+l \frac{\partial
M}{\partial l}\Big)\Big(-\frac{\partial^2 M}{\partial S^2 }dS^2
+\frac{\partial^2 M}{\partial Q^2 }dQ^2+2\frac{\partial^2
M}{\partial Q \partial l }dQ dl+\frac{\partial^2 M}{\partial l^2
}dl^2\Big) \,.\eea

Inserting here the expression for the mass (\ref{equ5}), we obtain
a rather cumbersome metric which cannot be written in a compact
form. Its corresponding scalar curvature turns out to be
nonzero, i.e., according with GTD there is thermodynamic
interaction. The explicit form of the curvature scalar cannot be
written in a compact form but we can see its behaviour in figure 12.

\begin{center}
\begin{figure}[h]
{\includegraphics[width=10cm]{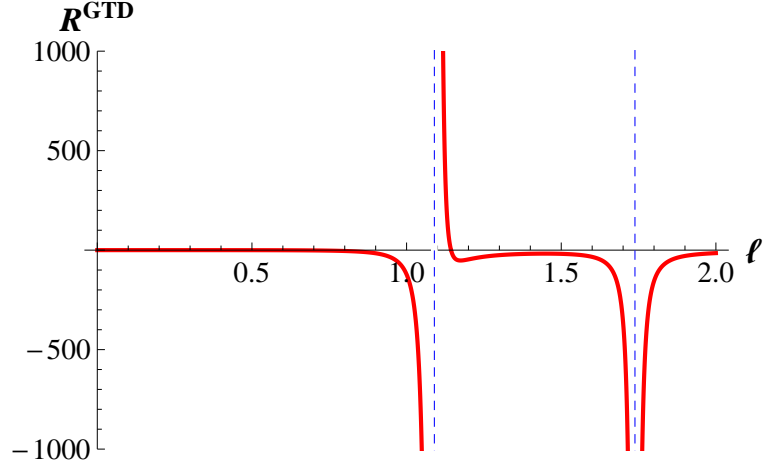}  \caption{The
curvature scalar $R^{GTD} $ as a function of $l$, with $Q=1$,
$S=10$, $s=5/2$ and $n=4$.}}\end{figure}
\end{center}

From figure 12 we can see that $R^{\text{GTD}}$ has two singularities but one of them corresponds to a point where the factor $f$ of the metric $g$ is zero,
$f=S \frac{\partial M}{\partial S} + Q \frac{\partial M}{\partial
Q}+l \frac{\partial M}{\partial l}=0$. This can  be seen in figure
13. This singular point is a singular point in the metric therefore it doesn't represent a physical divergence as the second singularity which takes place where the capacity becomes
infinite. This second singularity describes a second order phase transition.
Therefore, our analysis shows that the curvature singularities
reproduce the structure of the phase transitions of a black hole
with PMI source when we consider the cosmological constant as a
thermodynamic variable.

\begin{center}
\begin{figure}[h]
{\includegraphics[width=10cm]{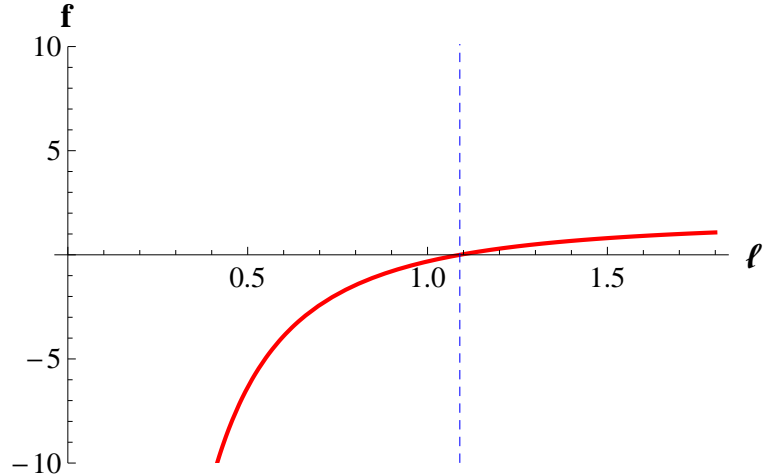} \caption{The factor $f=
S \frac{\partial M}{\partial S} + Q \frac{\partial M}{\partial
Q}+l \frac{\partial M}{\partial l}$ as a function of $l$, with
$Q=1$, $S=10$, $s=5/2$ and $n=4$.}}\end{figure}
\end{center}

\newpage
\section{Conclusions}
\label{conclusions}

In this paper, we analyzed the geometric structure of the
equilibrium manifold of a black hole with PMI source. By means of the
GTD formalism, which describes thermodynamic properties in terms
of differential geometry in a Legendre invariant way, we derive
the critical points that follow from the analysis of the
divergences of the heat capacity. In black hole thermodynamics,
the critical points of the heat capacity are usually associated
with the occurrence of second order phase transitions. Here we
analyzed the divergences of heat capacity and showed that GTD
reproduces the behaviour of these critical points. We also studied
the case $s=1$ corresponding to the
Reissner-Nordstr$\ddot{\mathrm{o}}$m and found that the PMI theory
admits second phase transition which are situated at those points
where the heat capacity diverges.

We analyzed the thermodynamic geometry based on the Weinhold
metric and found that it is a curved manifold for the black hole with PMI source
and the corresponding curvature diverges at some points, but these
points are not the ones at which the heat capacity for fixed
charge diverges. We conclude that Weinhold geometry does not
describe correctly the thermodynamic geometry for the black hole
with PMI source or at least it is not possible to give an interpretation of this divergences. We interpret this result as the need to impose
the Legendre invariance in the context of the geometric analysis
of the thermodynamics, since the Weinhold  metric are not
invariant with respect to Legendre transformations.

In the case where we consider the cosmological constant as a
thermodynamic variable, we found that a curvature singularity
appears exactly at that point where the phase transition
occurs. Thus, GTD reproduces correctly the thermodynamic phase
transition structure of a black hole with PMI source. These result
reinforces the conclusion that GTD is able to correctly reproduce
the phase transition structure of black holes.

\section*{Acknowledgements}

This work was supported by Conacyt-Mexico, Grant No. 166391 and
DGAPA-UNAM, Grant No. 106110. We would like to thank the members
of the GTD-group for fruitful comments and discussions. GA is
thankful to CONACYT, Grant No. 166391 Postdoctoral Fellowship.

\end{document}